\documentclass[apl,twocolumn,showpacs,superscriptaddress,letterpaper]{revtex4-1}
\usepackage{graphicx,amsmath,amsfonts}
\usepackage{tabularx}
\usepackage{bm}
\usepackage{mathrsfs}
\usepackage{appendix}
\usepackage{bbm}
\usepackage{color}
\def \vk {{\bf k}}
\def \vq {{\bf q}}
\def \vS {{\bf S}}

\def \mb {\mu_{\rm B}}

\def \vn {{\bf n}}
\def \vz {{\bf z}}

\def \vr {{\bf r}}
\def \vp {{\bf p}}
\def \vG {{\bf G}}

\def \lk {{\hat{l}_{z\vk }}}
\def \lok {{\hat{\overline l}_{z\vk }}}

\def \kov {\overline{k}}

\def \bmu {{\boldsymbol{\mu}}}

\begin{document}

\title{Gauge-Invariant Measure of the Magnon Orbital Angular Momentum}

\author{Randy S. Fishman}
\email{fishmanrs@ornl.gov
\newline
This manuscript has been authored in part by UT-Battelle, LLC, under contract DE-AC05-00OR22725 with the US Department of Energy (DOE). The US government retains and the publisher, by accepting the article for publication, acknowledges that the US government retains a nonexclusive, paid-up, irrevocable, worldwide license to publish or reproduce the published form of this manuscript, or allow others to do so, for US government purposes. DOE will provide public access to these results of federally sponsored research in accordance with the DOE Public Access Plan (http://energy.gov/downloads/doe-public-access-plan)}

\affiliation{Materials Science and Technology Division, Oak Ridge National Laboratory, Oak Ridge, Tennessee 37831, USA}

\date{\today}

\begin{abstract}
	Unlike the Berry phase, the orbital angular momentum (OAM) of magnons with two-dimensional wavevector $\vk $ in band $n$ 
	is not gauge invariant for arbitrary phase $\lambda_n(\vk)$ and so is not physically observable.
	However, by integrating the OAM over the orientation $\phi $ of wavevector $\vk $, we construct a gauge-invariant function $F_n(k)$.  
	Like $F_n(k)$, the average OAM for magnon band $n$ in a circle of radius $k$ is also gauge invariant and can be directly observed.  
	We demonstrate these results for a ferromagnet on a honeycomb lattice with Dzyalloshinskii-Moriya interactions between next-nearest neighbor spins. 
	With wavevectors $\vk $ restricted to the first Brillouin zone, the angular averaged OAM $F_n(k)$ then has opposite signs for lower and upper bands $n=1$ and 2 for all $k$.
	\end{abstract}

\keywords{spin-waves, orbital angular momentum}

\maketitle

\section{Introduction}

Magnons are quanta of spin excitations that can carry energy and information without incurring Joule heating.  In some magnetic materials, magnons may travel over 
centimeter distances \cite{Buttner2000} before appreciable decay.  Consequently, magnons have attracted a great deal of interest in the field of spintronics as replacements
for electrons.  While magnons have already exhibited a great many phenomenon of both scientific and technological interest such as the 
magnon thermal Hall \cite{Hirschberger15a, Hirschberger15b, Murakami17, Neumann22}, spin Seebeck \cite{Uchida10, Wu16}, and spin Nernst effects \cite{Cheng16, Zyuzin16, Zhang22}, the
field of ``magnonics" \cite{Wang21, Chumak22, Sheka22} has yet to reach full maturity.  

Due to spin-orbit (SO) coupling, the spin Hall effect produces a spin current perpendicular to a charge current \cite{Hirsch99, Zhang00}.  
Prior to its original observation by Onose {\it et al.} \cite{Onose2010}, the magnon Hall effect was 
predicted by Katsura {\it et al.} \cite{Katsura2010} using a Kubo formula that employed the Berry 
phase of a magnetic Hamiltonian.  A magnon wavepacket with center of mass at position $\vr_c$ obeys the 
semiclassical equation of motion \cite{xiao10}
\begin{equation}
\frac{d\vr_c}{dt }= \frac{\partial \omega_n(\vk )}{\partial \vk }-\frac{d \vk }{ dt} \times {\bf \Omega}_n(\vk ),
\end{equation}
where $\omega_n(\vk )$ is the magnon frequency, ${\bf \Omega}_n(\vk )$ is the Berry phase for magnon band $n$ and wavevector $\vk $, and
times is the cross product.
This relation predicts the bending of the magnon wavepacket in the presence of Dzyalloshinskii-Moriya (DM) interactions created by
SO coupling, i.e. the magnon Hall effect.  Like DM interactions, dipole interactions can also produce a nonzero Berry phase in ferromagnets (FMs) \cite{Okamoto17}.
The effects of the Berry phase induced by geometry on FMs in wires, ribbons, and spheres can be traced back to anisotropy and DM interaction energies \cite{Sheka22}. 
For a FM in the absence of DM or dipole interactions, both the Berry phase and the magnon Hall effect vanish.  

Like Ref.\,\cite{Katsura2010}, 
most subsequent work on magnon dynamics borrowed heavily from the semiclassical theory of electronic
band structure, with the Berry phase taking a central role \cite{xiao10}.  
Due to its roots in electronic band structure, the magnetic Berry phase is usually expressed in terms of Bloch functions $\vert u_n(\vk )\rangle $
as
\begin{equation}
{\bf \Omega}_n (\vk)=\frac{i}{2 \pi} 
\biggl\{ \frac{\partial }{\partial \vk } \times
\langle u_n(\vk ) \vert \frac{\partial }{\partial \vk }\vert u_n(\vk) \rangle \biggr\}.
\label{EqBerry}
\end{equation}
Earlier work on the Berry phase also specialized to FMs, so that the kinetic energy of a magnon can be written
as $-(\hbar k)^2/2m^*$, in analogy
with the electron kinetic energy,
where $m^*$ is the effective mass of the magnon. Parameterized in terms of $m^*$,
results for the magnon thermal Hall conductivity and other transport properties are valid only at low energies and temperatures,
where the dispersion of the FM magnon frequency $\omega_n(\vk )$ is quadratic.

Just as spin angular momentum underlays the magnetic interactions between moments, orbital angular momentum (OAM) underlays the Berry phase.
From a purely formal perspective, Matusmoto and Murakami \cite{Mat11a, Mat11b} described the OAM as the 
``self rotation" of the magnon wavepacket.  But from a physical point of view, the OAM of magnons has taken a decidedly secondary role to the Berry
phase in earlier work.

Considering the importance of OAM in other fields such as optics \cite{Shen19, Fang21, Rosen22, Watzel22} 
and electronic ``orbitronics" \cite{Go21, Han22, Dowinton22},
it is indeed surprising that more effort has not been made to understand the effects of OAM in thin film magnets.
We expect that the OAM of magnons will play important roles in information storage, communications technology, 
and in the coupling between magnons and other particles that can carry OAM, like electrons \cite{Mendis22}, phonons \cite{Zhang14, Hamada18}, and photons \cite{Marrucci06}.
In particular, the interaction between the spin and OAM of magnons might be utilized to control the flow and lifetime of magnetic excitations.

In the course of developing a quantum treatment for the magnon OAM, recent work \cite{Fishman22, Fishman23} provided four examples of 
collinear magnets where OAM appears when the exchange interactions 
create a non-Bravais lattice that violates
inversion symmetry and channels the magnon motion in nontrivial ways.  Two FMs and two antiferromagnets (AFs) were studied on zig-zag square and honeycomb lattices, as sketched in Fig.\,1.

Nevertheless, Refs.\,\cite{Fishman22} and \cite{Fishman23} side-stepped the important issue of gauge invariance \cite{Fukui2005}.
While the Berry phase is invariant under the gauge transformation 
\begin{equation}
\vert u_n(\vk )\rangle \rightarrow \vert u_n(\vk )\rangle \, e^{-i\lambda_n (\vk )}
\label{btr}
\end{equation}
for an arbitrary phase $\lambda_n (\vk )$, the OAM is {\it not} gauge invariant.  Quantities that 
depend on gauge are not considered to be physically observable \cite{xiao10}.  The absence of gauge invariance has stymied 
previous investigators and stalled earlier studies of the OAM.
In this paper, we show that a gauge-invariant, physically measurable function $F_n(k)$ can be obtained by 
integrating the OAM over the orientation $\phi $ of the wavevector $\vk =(k, \phi )$.

This paper is divided into five sections.  In Section II, we review some of the formal development originally presented 
in Ref.\,\cite{Fishman22}, now extended by including further 
quantum effects and simplified for collinear magnets.  Section III contains a derivation of the gauge invariant function $F_n(k)$. 
Section IV applies that function to the four case studies of Ref.\,\cite{Fishman22}.  The function $F_n(k)$ is nonzero only 
for a FM honeycomb lattice in the presence of DM interactions.  Section V explains 
how to translate expressions between the semiclassical and quantum languages.  A discussion is contained in Section VI.

\section{Quantum Formalism}

As explained in Ref.\,\cite{Fishman23}, the classical equations of motion \cite{Tsukernik66, Garmatyuk68} for the dynamical 
magnetization ${\bmu}_i=2\mb \,\delta \vS_i$ of a collinear magnet at site $i$ produce 
the linear momentum ${\bf p}_i$ \cite{Landau60}:
\begin{equation}
p_{i\alpha }=\frac{1}{4\mb M_0} (\bmu_i \times \vn_i ) \cdot \frac{\partial \bmu_i}{\partial x_{\alpha }},
\end{equation}
where $M_0=2\mb S$ is the static magnetization for a spin ${\bf{S}}_i$ pointing along $\vn_i=\pm \vz$.   
Using the $1/S$ quantization conditions
${\overline{\mu}_i}^+=\mu_{ix}n_{iz}+i\mu_{iy}=2\mb \sqrt{2S\hbar }\,a_i$ and ${\overline{\mu}_i}^-=\mu_{ix}n_{iz}-i\mu_{iy}=2\mb \sqrt{2S\hbar }\,a_i^{\dagger }$ for the dynamical magnetization
in terms of the local Boson operators $a_i$ and $a_i^{\dagger}$ satisfying the momentum-space
commutation relations $[a_{\vk }^{(r)}, a_{\vk'}^{(s)\dagger }]=\delta_{rs}\delta_{\vk,\vk'}$
and $[a_{\vk }^{(r)}, a_{\vk'}^{(s)}]=0$,
the quantized linear and OAM are given by
\begin{equation}
{\bf{p}}=-\frac{\hbar}{2}\sum_{r=1}^{M} {\sum_{\vk  }}' \vk \Bigl\{ a_{\vk }^{(r) \dagger } a_{\vk }^{(r)}+a_{\vk }^{(r)} a_{\vk }^{(r)\dagger }\Bigr\},
\end{equation}
\begin{eqnarray}
{\cal L}_z &= &\sum_i (\vr_i \times \vp_i )\cdot \vz \nonumber \\
&=&\frac{\hbar}{2}\sum_{r=1}^M {\sum_{\vk }}' \Bigl\{ a_{\vk}^{(r)}\,\lk \, a_{\vk }^{(r)\dagger } -a_{\vk }^{(r)\dagger }\, \lk \, a_{\vk }^{(r)}\Bigr\},
\end{eqnarray}
where $r$ and $s$ refer to the $M$ sites in the magnetic unit cell and 
\begin{equation}
\lk = -i \biggl( k_x\frac{\partial }{\partial k_y}-k_y\frac{\partial}{\partial k_x}\biggr)
\label{lk}
\end{equation}
is the OAM operator.  The prime restricts the sum over wavevectors $\vk $ to the first BZ of the magnetic 
unit cell.  The first relation specifies how the linear momentum ${\bf p}$, which can take any value inside or outside the first BZ, is expressed in terms
of wavevectors $\vk $ defined solely within the first BZ.

In terms of the  $a_{\vk }^{(r)}$ and $a_{\vk }^{(r)\dagger}$ operators, the second-order Hamiltonian $H_2$ can be written as
\begin{equation}
H_2={\sum_\vk }' {\bf v}_{\vk}^{\dagger }\cdot \underline{{\it L}}(\vk )\cdot {\bf v}_{\vk },
\label{defL}
\end{equation}
where the vector operators
\begin{equation}
{\bf v}_{\vk } =(a_{\vk }^{(1)},a_{\vk }^{(2)}\ldots  a_{\vk }^{(M)},a_{-\vk}^{(1)\dagger },a_{-\vk }^{(2)\dagger }\ldots a_{-\vk }^{(M)\dagger })
\end{equation}
satisfy 
$[{\bf v}_{\vk },{\bf v}^{\dagger }_{\vk'}] =\underline{N}\,\delta_{\vk ,\vk'}$
with
\begin{equation}
\underline{N} =
\left(
\begin{array}{cc}
\underline{I} &0 \\
0 & -\underline{I} \\
\end{array} \right)
\label{defn}
\end{equation}
and $\underline{I}$ is the $M$-dimensional identity matrix.  

\begin{figure}
\begin{center}
\includegraphics[width=8cm]{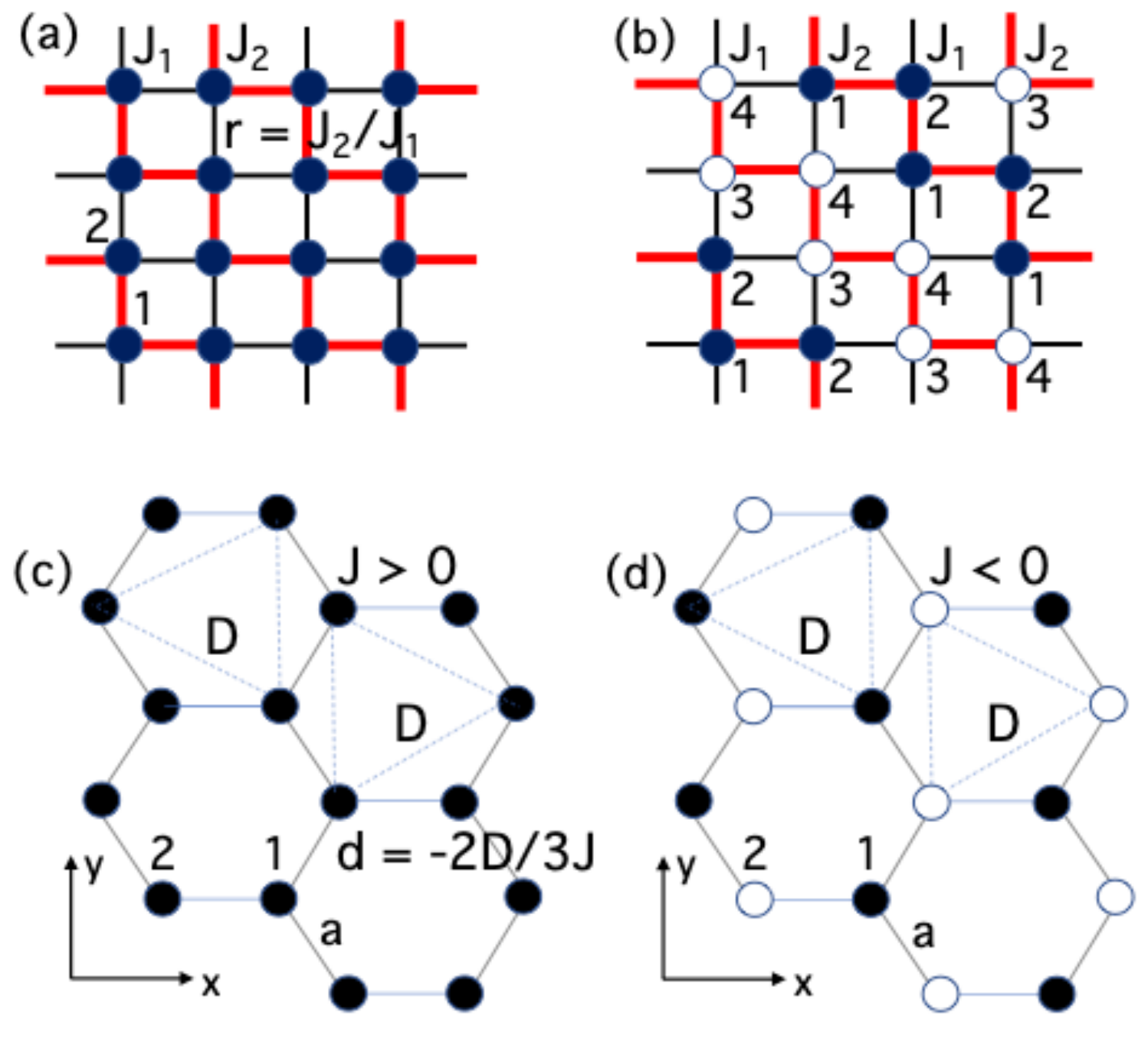}
\end{center}
\caption{Four case studies:  (a) a zig-zag lattice with FM interactions $J_1>0$, $J_2>0$, and ratio $r=J_2/J_1$, (b) a zig-zag lattice with AF interaction $J_1<0$ and FM interaction $J_2>0$, 
(c) a honeycomb lattice with FM interaction $J>0$ and DM interaction $D$ between next-nearest neighbors with $d=-2D/3J$, and (d) a honeycomb lattice with AF interaction $J<0$ and DM interaction $D$.  
In all cases, up spins are solid circles and down spins are empty circles.}
\label{Fig1}
\end{figure}

We then 
transform to the interacting vector operators 
\begin{equation}
{\bf w}_{\vk } =(b_{\vk }^{(1)},b_{\vk }^{(2)}\ldots  b_{\vk }^{(M)},b_{-\vk}^{(1)\dagger },b_{-\vk }^{(2)\dagger }\ldots b_{-\vk }^{(M)\dagger }),
\end{equation}
which satisfy
$[{\bf w}_{\vk },{\bf w}^{\dagger }_{\vk'}] =\underline{N}\,\delta_{\vk ,\vk'}$.
The relation between ${\bf v}_{\vk }$ and ${\bf w}_{\vk }$ is given by 
${\bf v}_{\vk }={\underline X}^{-1}(\vk )\cdot {\bf w}_{\vk }$, which may be expanded as
\begin{eqnarray}
a_{\vk }^{(r)}&=&\sum_n \Bigl\{ X^{-1}(\vk )_{rn}\,b_{\vk}^{(n)} +X^{-1}(\vk )_{r,n+M}\,b_{-\vk}^{(n)\dagger }\Bigr\} , \\
a_{-\vk }^{(r)\dagger }&=&\sum_n \Bigl\{ X^{-1}(\vk )_{r+M,n}\,b_{\vk}^{(n)} +X^{-1}(\vk )_{r+M,n+M}\,b_{-\vk}^{(n)\dagger }\Bigr\}.
\nonumber
\end{eqnarray}
The matrix $\underline{X}^{-1}(\vk )$ obeys the eigenvalue equation \cite{fishmanbook18}
\begin{equation}
\underline{\Lambda} (\vk )\cdot\underline{X}^{-1}(\vk ) = \epsilon_n(\vk )\,\underline{X}^{-1 }(\vk ),
\label{egv}
\end{equation}
where ${\underline \Lambda}(\vk ) = {\underline N}\cdot \underline{L}(\vk )$
and $\epsilon_n(\vk )=\hbar \omega_n(\vk )/2$ ($n=0,\ldots ,M$) or $-\hbar \omega_n(-\vk )/2$ ($n=M+1,\ldots ,2M$).
This expression is the quantum analogue of the semiclassical relation 
\begin{equation}
H_2\vert u_n(\vk )\rangle =\hbar \omega_n(\vk )\vert u_n (\vk )\rangle .
\label{bhe}
\end{equation}
Hence, $X^{-1 }(\vk )_{rn}$ can be considered the $n$th eigenfunction of the $2M \times 2M$ magnon energy matrix ${\underline{\Lambda}}(\vk )$.

In terms of the interacting Boson operators, we find 
\begin{equation}
{\bf{p}}=-\frac{\hbar}{2}\, {\sum_{n, \vk  }}' \vk \Bigl\{ 2b_{\vk }^{(n) \dagger } b_{\vk }^{(n)}+1 \Bigr\},
\end{equation}
\begin{equation}
{\cal L}_z = {\sum_{n ,\vk }}' O_n(\vk )\Bigl\{ 2b_{\vk }^{(n) \dagger } b_{\vk }^{(n)}+1 \Bigr\}
+ ({\bf r}_c \times {\bf p}_c)\cdot \vz,
\label{Lzbf}
\end{equation}
where
\begin{eqnarray}
&&O_n(\vk )= \frac{\hbar}{2}\sum_{r=1}^M\Bigl\{ 
X^{-1}(\vk )_{rn}\,\lk \, X^{-1}(\vk )_{rn}^* \nonumber \\
&&- X^{-1}(\vk)_{r+M,n}\,\lk \, X^{-1}(\vk )^*_{r+M,n}\Bigr\}.
\label{Lzdef}
\end{eqnarray}
The last contribution $({\bf r}_c \times {\bf p}_c)$ to ${\cal L}_z$ is the OAM of the center-of-mass of the magnon 
wavepacket, as defined by Chang and Niu \cite{Chang96}.

Because the factor 
$2b_{\vk }^{(n) \dagger } b_{\vk }^{(n)}$ was treated incorrectly, Ref.\,\cite{Fishman22} undercounted
the OAM ${\cal L}_z$ for mode $n$ by a factor of three \cite{Fishman23E}.
Of course, $2\langle b_{\vk }^{(n) \dagger } b_{\vk }^{(n)}\rangle = 2$ for a single magnon with 
wavevector $\vk $ in state $n$.
For collinear spin states on a non-centrosymmetric lattice {\it without} DM or dipole interactions, time-reversal symmetry requires \cite{Go21}
$\underline{X}^{-1}(-\vk )=\underline{X}^{-1}(\vk )^*$ so that
$O_n(\vk )=-O_n(-\vk )$ is an odd function of $\vk $.   It follows that 
the OAM for a given magnon band $n$ would vanish when integrated over a ring of radius $k$ 
within the two-dimensional BZ.

Using the semiclassical notation, the OAM can be written as
\begin{equation}
O_n (\vk) =-\frac{i \hbar}{2} \biggl\{\vk \times \langle u_n(\vk )\vert \frac{\partial }{\partial \vk }\vert  u_n(\vk ) \rangle \biggr\}\cdot \vz. 
\end{equation}
On the other hand,
the Berry phase ${\bf \Omega}_n(\vk )$
can be written in terms of the quantum eigenfunctions $\underline{X}^{-1}(\vk )$ as
\begin{eqnarray}
&&{\bf \Omega}_n (\vk )= \frac{i}{2\pi}\sum_{r=1}^M\biggl\{ 
\frac{\partial X^{-1}(\vk )_{rn}^*}{\partial \vk } \times \frac{\partial X^{-1}(\vk )_{rn}}{\partial \vk } \nonumber \\
&&- \frac{ \partial X^{-1}(\vk)_{r+M,n}^*}{\partial \vk} \times  \frac {\partial X^{-1}(\vk )_{r+M,n}}{\partial \vk } \biggr\}.
\label{Bfdef}
\end{eqnarray}
A guide to translating expressions between the semiclassical and quantum languages is provided in Section V.

\section{Gauge Invariance}

In the semiclassical language,
each Bloch function $\vert u_n(\vk )\rangle $ can be multiplied by an arbitrary phase factor $\exp (-i\lambda_n(\vk))$
as in Eq.\,(\ref{btr}).
In the quantum language, each eigenfunction $X^{-1 }(\vk )_{rn}$ can also be multiplied by an arbitrary phase factor 
so that
\begin{equation}
X^{-1}(\vk)_{rn }\rightarrow X^{-1}(\vk )_{rn}\,e^{-i\lambda_n(\vk )},
\label{gtr}
\end{equation}
where $\lambda_n(\vk )$ may depend on $\vk $ and band index $n$ but not on site $r$.  
Under a gauge transformation,
\begin{equation}
O_n(\vk )\rightarrow O_n(\vk )+\frac{\hbar }{2}\biggl( k_x \frac{\partial }{\partial k_y}-k_y \frac{\partial }{\partial k_x}\biggr)\lambda_n(\vk ) ,
\label{Lztr}
\end{equation}
\begin{equation}
{\bf \Omega}_n(\vk )\rightarrow {\bf \Omega}_n(\vk ),
\end{equation}
both of which use the normalization condition
$\underline{X}^{-1}(\vk )\cdot \underline{N} \cdot \underline{X}^{-1\, \dagger }(\vk )=\underline{N}$ (equivalent to 
$\langle u_n(\vk )\vert u_n(\vk )\rangle =1$), or
\begin{equation}
\sum_{r=1}^M \Bigl\{ \vert X^{-1}(\vk )_{rn}\vert^2 - \vert X^{-1}(\vk )_{r+M,n}\vert^2\Bigr\}=1.
\label{sumx}
\end{equation}
Whereas the Berry phase is invariant for any phase factor $\lambda_n(\vk )$, Eq.\,(\ref{Lztr}) indicates that the OAM is not.

After decomposing $\vk = (k, \phi )$ in terms of its magnitude $k$ and orientation $\phi$, we find that
\begin{equation}
\biggl( k_x \frac{\partial }{\partial k_y}-k_y \frac{\partial }{\partial k_x}\biggr)\lambda_n(\vk   ) = \frac{\partial }{\partial \phi  }\lambda_n(k,\phi ).
\label{trans}
\end{equation}
Due to the $\phi $ dependence of the phase $\lambda_n (\vk )=\lambda_n (k,\phi)$, $O_n(\vk )$ is {\it not} gauge invariant.
In order to obtain an observable measure of the OAM,
we construct the function
\begin{equation}
F_n(k)= \int_0^{2\pi }\frac{d\phi }{2\pi } \, O_n(\vk ).
\label{lama}
\end{equation}
Under a gauge transformation,
\begin{eqnarray}
F_n(k)&\rightarrow &F_n(k) +\frac{\hbar }{2} \int_0^{2\pi }\frac{d \phi }{2\pi } \, \frac{\partial }{\partial \phi  }\lambda_n(k,\phi )\nonumber \\
 &=& F_n(k) +\frac{\hbar }{2} \Bigl\{ \lambda_n(k,2\pi )-\lambda_n(k,0)\Bigr\}\nonumber \\
 &=& F_n(k),
 \label{lamb}
\end{eqnarray}
which assumes only that $\lambda_n(\vk )$ is a single-valued function of the wavevector $\vk $ \cite{stokes}.
Hence, $F_n(k)$ is a gauge-invariant function.  Of course, $F_n(k)$ is nonzero only for a band $n$ with a net OAM 
when integrated over a ring for all angles $\phi $ with a fixed $k$.   

To better understand the results for $O_n(\vk )$, we can also evaluate the OAM averaged over a circle of radius $k$:
\begin{eqnarray}
O_{n,\rm av}(k) &=& \frac{2}{k^2}\int_0^k dq\, q  F_n(q) \nonumber \\
&=& \frac{1}{\pi k^2 } \int d\vq \, O_n(\vq )\, H (k - q),
\end{eqnarray}
where the Heaviside function $H (x)$ is defined so that $H (x)=1$ for $x>0$ and 0 otherwise.  Like $F_n(k)$, $O_{n,\rm av}(k)$
is also gauge invariant.

\section{Case Studies}

Now consider the four examples sketched in Fig.\,1 and discussed in Ref.\,\cite{Fishman22}. 
Since DM and dipole interactions are absent for the FM and AF zig-zag lattices in Figs.\,1(a) and (b),
$O_n(\vk )=-O_n(-\vk)$ is odd in $\vk$.  This immediately implies that $F_n(k)=0$.  
For the AF honeycomb lattice in Fig.\,1(d), DM interactions shift the magnon frequencies but do not affect the magnon energy matrix 
$\underline{\Lambda }(\vk ) = \underline{N}\cdot \underline{L}(\vk )$ in any non-trivial way.  So once again $O_n(\vk )=-O_n(-\vk)$ and $F_n(k)=0$.

The only case that satisfies the condition $F_n(k)\ne 0$ is the FM honeycomb with $d=-2D/3J >0$ shown in Fig.\,1(c),
where $J>0$ is the nearest-neighbor exchange interaction and $D<0 $ is the next-nearest neighbor DM interaction, which breaks time-reversal symmetry.   
Strong easy-axis anisotropy
$-K\sum_i {S_{iz}}^2$ prevents the spins from tilting away from the $z$ axis.
The $4\times 4$ matrix $\underline{L}(\vk )$ defined by Eq.\,(\ref{defL}) can then be written
\begin{equation}
\underline{L}(\vk ) =
\frac{3JS}{2} \left(
\begin{array}{cccc}
1-G_{\vk }  & -\Gamma_{\vk}^* & 0 & 0 \\
-\Gamma_{\vk } & 1+G_{\vk }  & 0 & 0 \\
0 & 0 & 1+G_{\vk}  & - \Gamma_{\vk }^*\\
0 & 0 &-\Gamma_{\vk } & 1-G_{\vk} \\
\end{array} \right),
\end{equation}
where $G_{\vk }=d \,\Theta_{\vk }$,
\begin{equation}
\Theta_{\vk} = 4\cos(3k_xa/2) \sin(\sqrt{3} k_ya/2)-2\sin(\sqrt{3}k_ya),
\end{equation}
\begin{equation}
\Gamma_{\vk} =\frac{1}{3}\Bigl\{ e^{ik_xa}+2e^{-ik_xa/2} \cos(\sqrt{3}k_ya/2) \Bigr\}.
\end{equation}
The magnon mode energies are given by
\begin{equation}
\hbar \omega_1(\vk)=3JS ( 1 + \kappa - \eta_{\vk } )
\label{om1},
\end{equation}
 \begin{equation}
\hbar \omega_2(\vk)=3JS ( 1 + \kappa + \eta_{\vk } ),
\label{om2}
\end{equation}
with $\eta_{\vk }=\sqrt{\vert \Gamma_{\vk }\vert^2 +G_{\vk }^2}$.
Because the anisotropy $\kappa =2K/3\vert J\vert $ merely shifts the magnon energies $\hbar \omega_n(\vk )$
but does not affect the OAM, its contribution to $\underline{L}(\vk )$ is omitted.  

After some manipulations, we find 
\begin{eqnarray}
X^{-1}(\vk)_{11}&=&-\frac{1}{2c_1(\vk )\eta_{\vk }},\\
X^{-1}(\vk)_{12}&=&\frac{1}{2c_2(\vk )\eta_{\vk }},\\
X^{-1}(\vk)_{21}&=&\frac{\eta_{\vk }+G_{\vk}}{2c_1(\vk )\Gamma_{\vk }^* \,\eta_{\vk }},\\ 
X^{-1}(\vk)_{22}&=&\frac{\eta_{\vk } -G_{\vk}}{2c_2(\vk )\Gamma_{\vk }^* \,\eta_{\vk }},
\end{eqnarray}
while the 31, 32, 41, and 42 matrix elements of $\underline{X}^{-1}(\vk)$ vanish.
The normalization condition $\underline{X}^{-1}(\vk )\cdot \underline{N} \cdot \underline{X}^{-1\, \dagger }(\vk )=\underline{N}$ gives
\begin{equation}
c_1(\vk ) =\frac{e^{i\lambda_1(\vk )}}{\sqrt{2\eta_{\vk }(\eta_{\vk }-G_{\vk })}},
\label{defc1}
\end{equation}
\begin{equation}
c_2(\vk ) =\frac{e^{i\lambda_2(\vk )} }{\sqrt{2 \eta_{\vk } (\eta_{\vk }+G_{\vk })}}.
\label{defc2}
\end{equation}
Here, $\lambda_n(\vk )$ are arbitrary phases because the normalization conditions only determine the amplitudes $\vert c_n(\vk )\vert $.
Since Eq.\,(\ref{egv}) is a linear eigenvalue equation, the column vectors $X_{rn}^{-1}(\vk )$ for modes $n=1$ and 2 are only determined up to 
overall arbitrary phase factors $\exp (-i\lambda_n(\vk ))$.

Regardless of those phase factors, the Berry phase along $\vz$ is given by 
\begin{equation}
\Omega_{1z}(\vk ) =-i\frac{d }{4\pi } \, \frac{\Gamma_{\vk}^*}{\vert \Gamma_{\vk }\vert } 
\Biggl\{ \frac{\partial \Theta_{\vk } /\eta_{\vk } }{\partial \vk } \times \frac{\partial \Gamma_{\vk }/\vert \Gamma_{\vk } \vert }{\partial \vk }  \Biggr\} \cdot \vz ,
\label{bf3}
\end{equation}
which is plotted in Fig.\,2 for four different values of $d$.  As the above expression makes clear, 
the Berry phase vanishes for $d=0$.  Notice that the Berry phase is six-fold symmetric and always positive for mode 1.  The peaks of the Berry phase rotate
by 30$^o$ when $d$ exceeds about 0.06.  At that value for $d$, the maximum amplitude of the Berry phase reaches a minimum.
For mode 2, $\Omega_{2z}(\vk )=-\Omega _{1z}(\vk )<0$.

\begin{figure}
\begin{center}
\includegraphics[width=8.5cm]{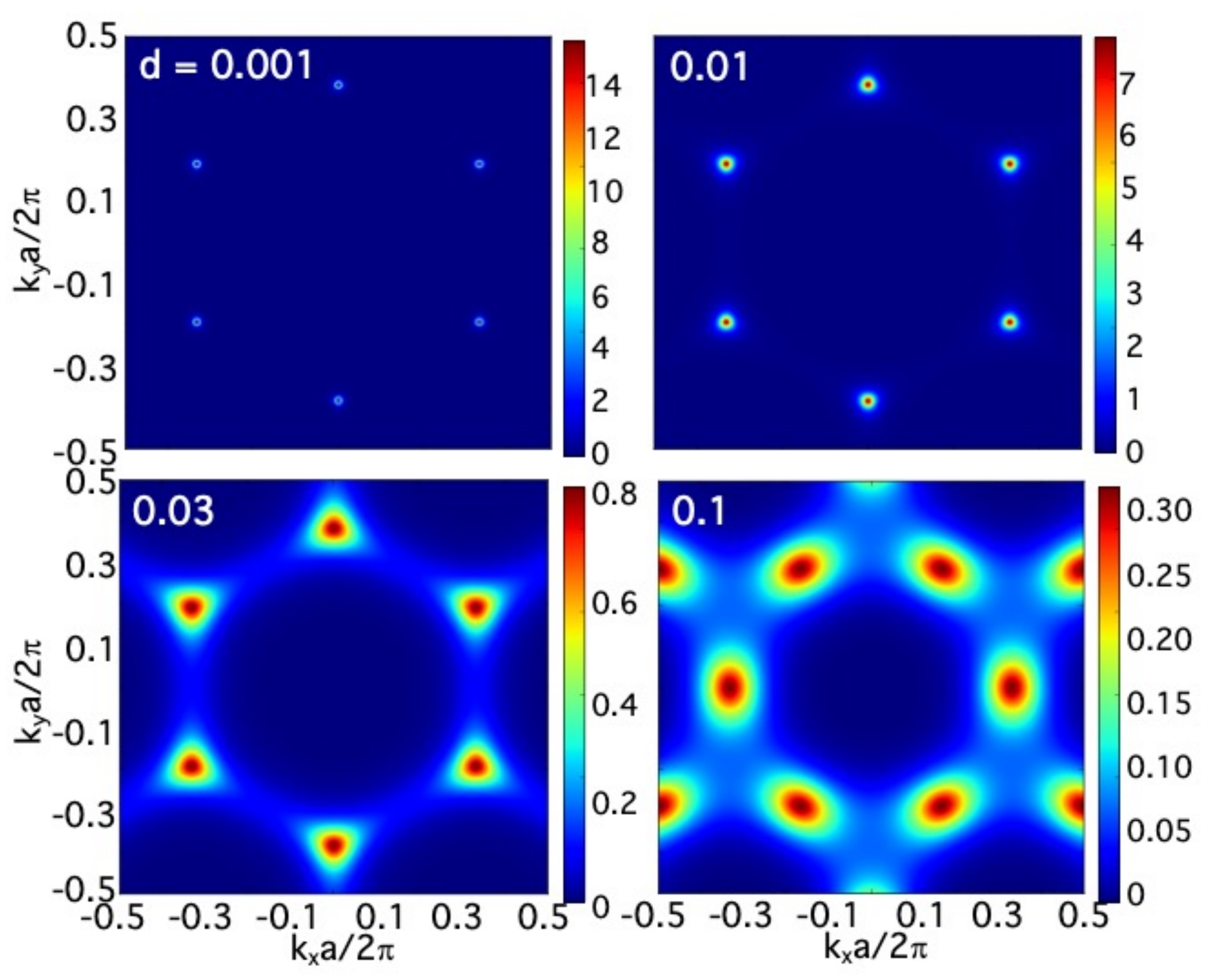}
\end{center}
\caption{Berry phase $\Omega_{1z}(\vk )/\hbar $ for a honeycomb lattice with FM exchange $J> 0$ between neighboring up spins and DM interaction $D$
between next-neighbor as shown in Fig.\,1(c) for values of $d=-2D/3J$ between 0.001 and $0.1$, for band 1.}
\label{Fig2}
\end{figure}

Since the DM interactions break time-reversal symmetry,
$O_n(\vk )$ contains 
both even and odd terms with respect to $\vk $ due to the $G_{\vk }=-G_{-\vk} \sim d$ functions in $\underline{X}^{-1}(\vk )$.
Assuming that $c_1(\vk )$ and $c_2(\vk )$ are both real or that $\lambda_n(\vk )=0$, the OAM for the lower and upper bands are given by 
\begin {equation}
O_1(\vk )=\frac{\hbar }{4} \biggl\{ 1 + \frac{d\,\Theta_{\vk } }{\eta_{\vk }} \biggr\}  \frac{\Gamma_{\vk}}{ \vert \Gamma_{\vk }\vert }\,
 \lk \,  \frac{\Gamma_{\vk }^*}{\vert \Gamma_{\vk }\vert },
 \label{lzl}
\end{equation}
\begin {equation}
O_2 (\vk)=\frac{\hbar }{4} \biggl\{ 1 -\frac{d\,\Theta_{\vk } }{\eta_{\vk }} \biggr\}  \frac{\Gamma_{\vk}}{ \vert \Gamma_{\vk }\vert }\,
 \lk \,  \frac{\Gamma_{\vk }^*}{\vert \Gamma_{\vk }\vert }.
 \label{lzu}
 \end{equation}
We plot the OAM versus $\vk $ for the upper and lower magnon bands in the top and lower panels, 
respectively, of Fig.\,3.  The OAM for the two bands are identical for the degenerate bands when $d=0$ but they differ for the 
non-degenerate bands when $d>0$.  As seen for the upper or lower bands with $d=0$, the OAM peaks at the boundaries of the 
repeated first BZ of the honeycomb lattice.  While the OAM of the panels with $d=0$ obey odd symmetry $O_n(-\vk )=-O_n(\vk)$,
the OAM of the panels with $d>0$ violate this symmetry.  The first BZ of the magnetic unit cell is the solid hexagon
drawn on the bottom left panel of Fig.\,3.  
 
Surprisingly, the results of Fig.\,3 are very different than those presented in Refs.\,\cite{Fishman22} and \cite{Fishman23},
where the linear terms $k_{\alpha }$ in the OAM operator $\lk $ of Eq.\,(\ref{lk}) were replaced by periodic 
functions ${\bar k}_{\alpha }$, 
\begin{eqnarray}
\kov_xa&=&\sin(3k_xa/2)\cos(\sqrt{3}k_ya/2), \\
\kov_ya&=& \frac{1}{\sqrt{3}}\Bigl\{ \sin(\sqrt{3}k_ya/2)\cos(3k_xa/2) \nonumber \\
&&+\sin(\sqrt{3}k_ya)\Bigr\},
\end{eqnarray}
constructed so that ${\bar k}_{\alpha }(\vk + \vG_m )={\bar k}_{\alpha }(\vk )$
for any reciprocal lattice vector $\vG_m$ of the honeycomb lattice.  With the periodic
OAM operator
\begin{equation}
\lok = -i \biggl( \kov_x\frac{\partial }{\partial k_y}-\kov_y\frac{\partial}{\partial k_x}\biggr),
\label{lok}
\end{equation}
the OAM ${\overline O}_n(\vk )$ also becomes a periodic function of $\vk $.
By contrast, the OAM plotted in Fig.\,3 is clearly not a periodic function.  In particular, the 
OAM $O_n(\vk )$ increases in size with the magnitude of $k$.
With the periodic $\kov_x$ and $\kov_y$ defined above, Eq.\,(\ref{trans}) no longer holds and a gauge-invariant, angular-averaged OAM
cannot be derived.

\begin{figure}
\begin{center}
\includegraphics[width=8.5cm]{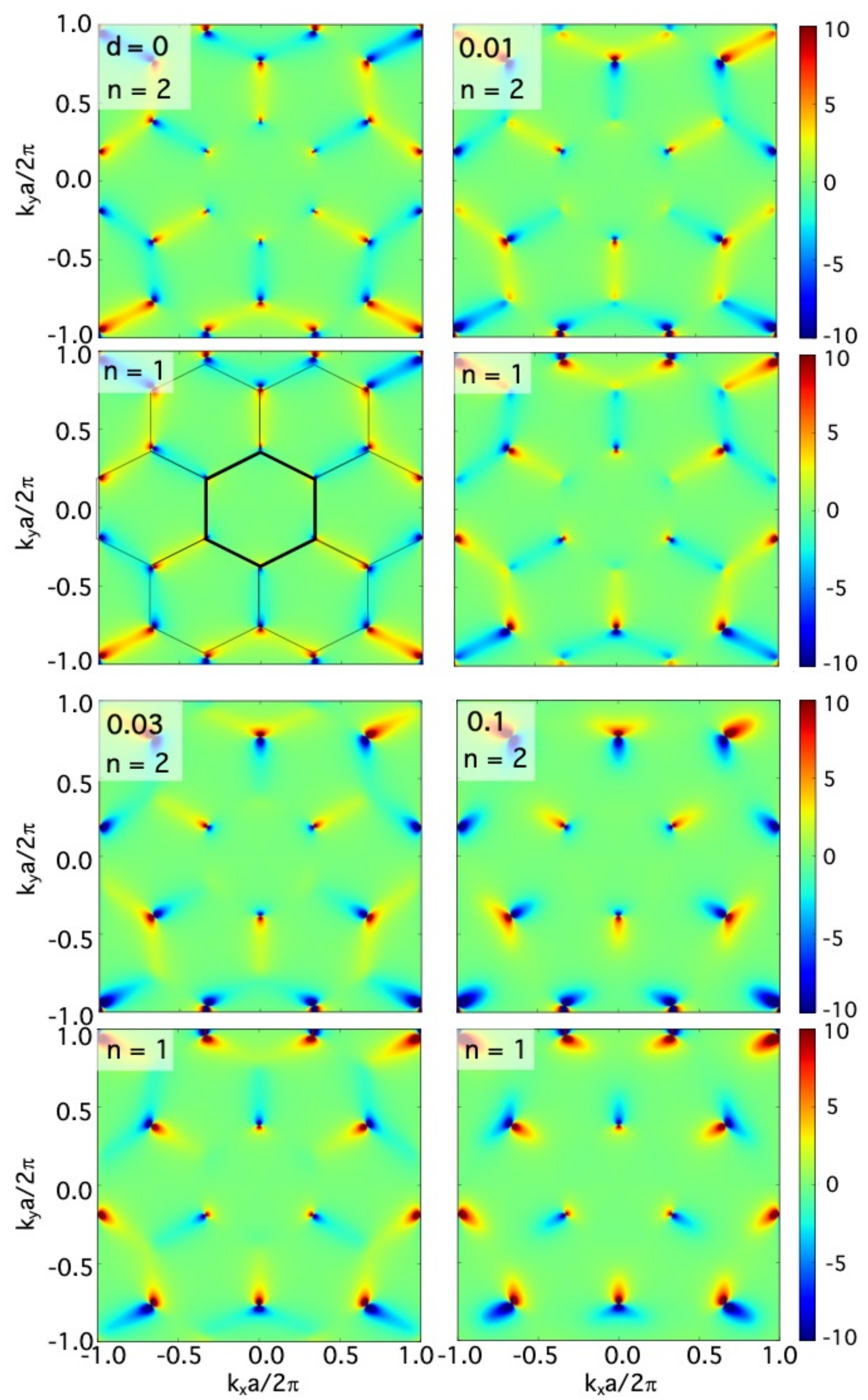}
\end{center}
\caption{A honeycomb lattice with FM exchange $J>0$ between neighboring up spins
and DM interaction $D$ between next-nearest
neighbors as shown in Fig.\,1(c).  The OAM $O_n (\vk )/\hbar $ for the upper, $n=2$ (top) and lower, $n=1$ (bottom) bands versus $\vk $ for different values of $d=-2D/3J$.  The repeated first
BZ boundary of the magnetic unit cell (thick solid hexagon) is sketched on the bottom panel for $d=0$.}
\label{Fig3}
\end{figure}

For magnon band 1, the gauge-invariant quantity $F_1(k)$ is plotted in 
Fig.\,4(a).  For band 2, $F_2(k)=-F_1(k)$, as seen from Fig.\,3.  With increasing $k$,
$F_1(k)$ oscillates between positive and negative values and is marked by sharp kinks at its maxima and minima.  
The positions $k$ for the maxima in $|F_n(k)|$ in Fig.\,4(a) correspond to the corners of the hexagons in the lower left panel of Fig.\,3, where 
red and blue regions meet.  Notice that $dF_1(k)/dk \ge 0$ while $dF_2(k)/dk \le 0$ with peaks in the derivatives $\vert dF_n(k)/dk\vert $
at the discontinuities of $F_n(k)$.  

We emphasize that the OAM must change for different choices of the complex phases $\lambda_n(\vk )$ in Eqs.\,(\ref{defc1}) and (\ref{defc2}).  
Hence, $O_n(\vk )$ plotted in Fig.\,3 are not themselves observable.
However, by integrating $\vk =(k,\phi )$ over all angles $\phi $ for a fixed $k$, we have resolved that phase ambiguity 
and created gauge-invariant, observable functions $F_n(k)$.

The functions $O_{1,{\rm av}}(k)$ are plotted in Fig.\,4(b) for four values of $d$ from 0.001 to 0.1.  We find that $O_{1, {\rm av}}(k)$ is an oscillatory function that contains cusps at 
positive peaks when $F_n(k)$ discontinuously drops and negative valleys when $F_n(k)$ rises through 0. 
The first such cusp lies at the corners of the first hexagonal BZ with $k=2\sqrt{3}/9(2\pi/a)\approx 0.385(2\pi/a)$.  For $d=0.1$, the average OAM peaks at $0.236 \hbar $ at that cusp.
Other cusps lie at the corners of the hexagons drawn in the lower left
panel of Fig.\,3.  As $k$ increases, the average OAM tends to zero.
By contrast, the average OAM $O_{n,\rm av}(k)$ of each band of the honeycomb lattice with imposed wavevector periodicity \cite{Fishman22} is nonzero (and opposite) as $k\rightarrow \infty $.
Note that $O_{2,{\rm av}}(k)=-O_{1,{\rm av}}(k)$ so that the net average of the two bands vanishes.  

\begin{figure}
\begin{center}
\includegraphics[width=7.5cm]{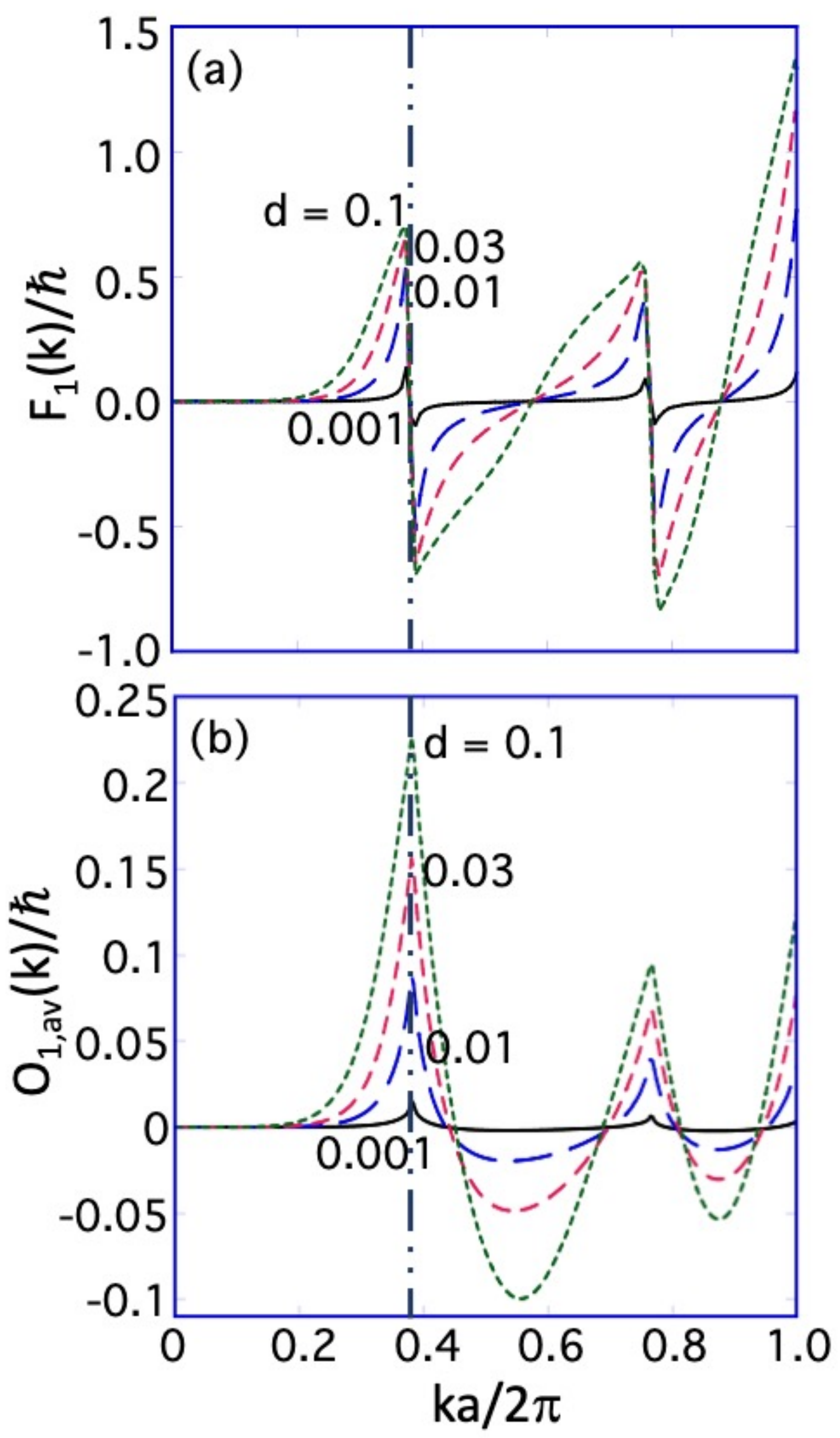}
\end{center}
\caption{(a) Gauge-invariant function $F_1(k)/\hbar $ for magnon band 1 of a FM open honeycomb lattice versus $k$ for four values of the 
DM interaction $d$.  (b) The average OAM $O_{1 ,{\rm av}}(k)/\hbar $ versus $ka/2\pi $ of band 1 for the same four values of 
$d$.  For magnon band 2, $F_2(k) =-F_1(k)$ and $O_{2 ,{\rm av}}(k) = -O_{1 ,{\rm av}}(k) $.  The dot-dash vertical line marks $k_{\rm max}$ for the 
first BZ in both (a) and (b).}
\label{Fig4}
\end{figure}

\section{Translating between the Semiclassical and Quantum Languages}

In order to reconcile Eqs.\,(\ref{EqBerry}) and (\ref{Bfdef}) for the Berry phase, we must
we must define 
\begin{equation}
\langle u_n(\vk )\vert {\hat A}\vert u_n(\vk )\rangle_{\rm cell} \equiv \sum_{r=1}^M \langle {\underline v}_{rn}(\vk ) \vert {\hat A}\,{\underline N}\vert {\underline v}_{rn} (\vk )\rangle ,
\label{t1}
\end{equation}
where
\begin{equation}
\vert {\underline v}_{rn }(\vk )\rangle =  { {X^{-1}_{rn}(\vk )} \choose {X^{-1}_{r+M,n}(\vk)} },
\end{equation}
\begin{equation}
\langle {\underline v}_{rn }(\vk )\vert =  \Big( {X^{-1}_{rn}(\vk )^*},  {X^{-1}_{r+M,n}(\vk)^*} \Big),
\end{equation}
and the integral over the magnetic unit cell on the $lhs$ of Eq.\,(\ref{t1}) is replaced by a sum over sites $r$ within the magnetic unit cell
on the $rhs$ of that expression.
As required, this transformation implies that 
\begin{eqnarray}
\langle u_n(\vk )\vert {\hat I}&&\vert u_n(\vk )\rangle \nonumber \\
&&=\sum_{r=1}^M \langle {\underline v}_{rn}(\vk ) \vert {\underline N}\vert {\underline v}_{rn} (\vk )\rangle \nonumber \\
&& = 
\sum_{r=1}^M \Bigl\{ \vert X^{-1}(\vk )_{rn}\vert^2 - \vert X^{-1}(\vk )_{r+M,n}\vert^2\Bigr\}\nonumber \\
&& =1,
\end{eqnarray}
which uses the normalization condition of Eq.\,(\ref{sumx}).  

\section{Discussion}

The function $F_n(k)$ gives the average OAM over a ring with wavevector $k$.  Because it is gauge invariant, $F_n(k)$ is also measureable.   
It yields the average OAM at $k=\vert \vk \vert$, but not the angles $\phi $ at which that OAM can be detected.  
One of the greatest barriers to studies of the OAM at wavevector $\vk $ has been its lack of gauge invariance.  
By integrating the OAM over the orientation $\phi $ of $\vk $, we have constructed gauge-invariant functions $F_n(k)$ and
$O_{n,\rm av}(k)$ that are nonzero for the FM honeycomb lattice but vanish for the AF honeycomb lattice and in the absence of DM and dipole interactions.  

\begin{figure}
\begin{center}
\includegraphics[width=8.5cm]{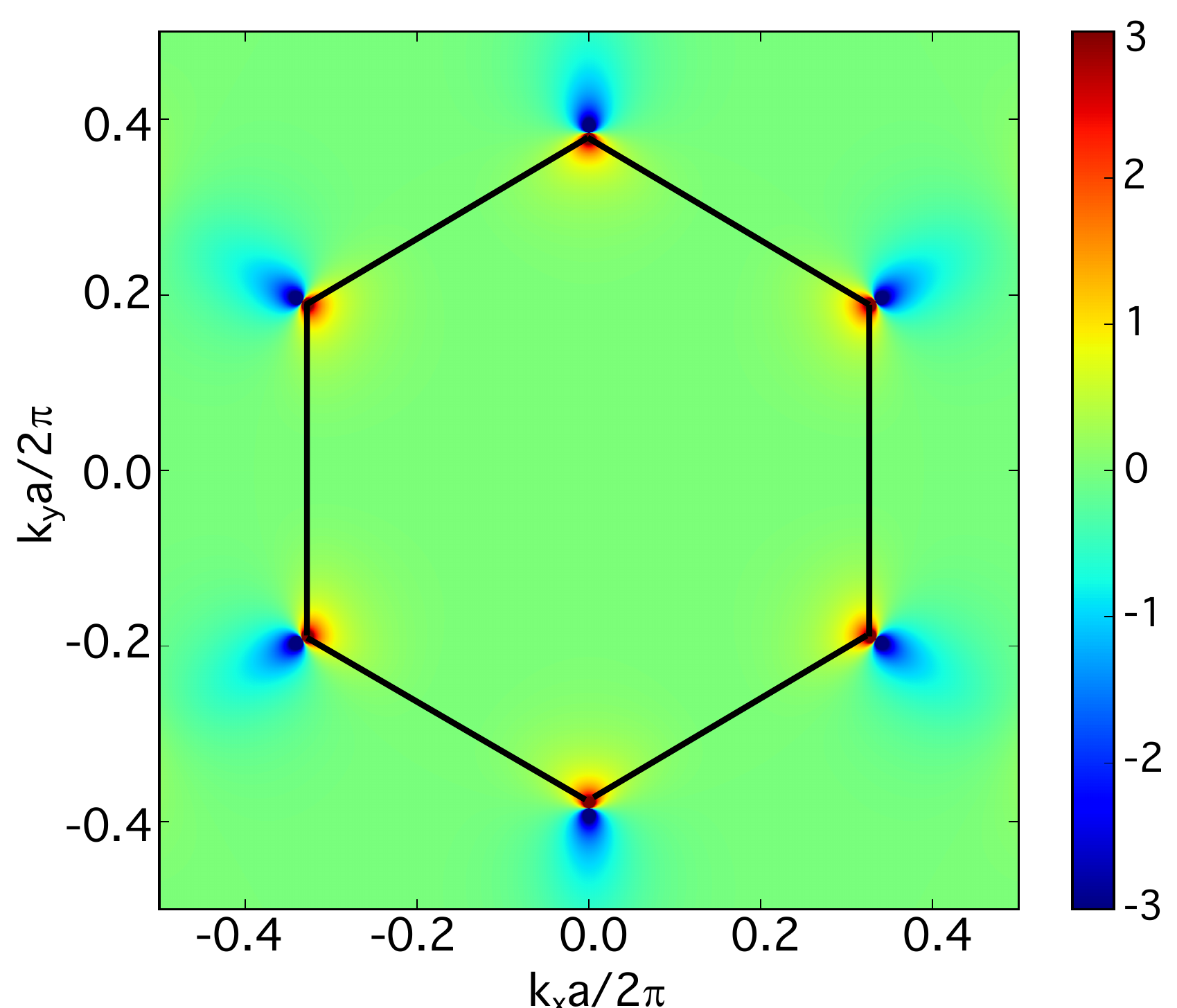}
\end{center}
\caption{The even OAM function $O_1'(\vk )$ in units of $\hbar $ constructed by subtracting the odd part of the OAM from $O_1(\vk )$ for the FM honeycomb lattice with $d=0.1$.
The first BZ is bordered by the solid hexagon.}
\label{Fig5}
\end{figure}

While $O_n(\vk )$ is not a periodic function of $\vk $, 
physical quantities like ${\cal L}_z$ in Eq.\,(\ref{Lzbf}) impose limits on the wavevector $\vk $ by restricting sums to the first BZ of the magnetic unit cell.  
For the FM honeycomb lattice, the magnetic unit cell is the hexagon sketched by the solid line in Fig.\,5.  
The maximum amplitude of the wavevector within the
first BZ lies at its corners with $k_{\rm max} = 2\sqrt{3}/9(2\pi/a) \approx 0.385 (2\pi )/a $.  

A periodic OAM can be constructed by first subtracting the non-physical, odd function $O_n^{\rm odd}(\vk )$ (obtained by neglecting DM and dipole interactions)
from $O_n(\vk )$ so that the remainder 
$O_n'(\vk )=O_n(\vk )-O_n^{\rm odd}(\vk )$ is an even function of $\vk $.  For the FM honeycomb lattice,
\begin {equation}
O_1'(\vk )=-O_2'(\vk )= \frac{\hbar }{4}\frac{d\,\Theta_{\vk } }{\eta_{\vk }}  \frac{\Gamma_{\vk}}{ \vert \Gamma_{\vk }\vert }\,
 \lk \,  \frac{\Gamma_{\vk }^*}{\vert \Gamma_{\vk }\vert },
 \label{elzl}
\end{equation}
which is proportional to $d$.  Like the Berry phase $\Omega_{1z}(\vk )$, $O_1'(\vk )$ plotted in Fig.\,5 is also a six-fold symmetric function of $\vk $.  
A periodic OAM can then be constructed by tiling $\vk $-space with the first BZ of $O_n'(\vk )$.

Nevertheless, it is doubtful that a magnon with wavevector outside the first BZ has any physical significance.  Certainly, any such magnon would rapidly decay via higher-order quantum
processes into single magnons within the first BZ while conserving energy, momentum, spin, and OAM.
Considering only magnons within the first BZ of the open honeycomb lattice, we can reach several conclusions.  From Fig.\,4, we see that both $F_1(k)$ and $O_{1,{\rm av}}(k)$ (band 1) are positive for all $k$ 
within the first BZ.  For $d=0.1$, band 1 will then have an average OAM of about $0.24\hbar $ while band 2 will have an average OAM of about $-0.24\hbar $.

A natural question is whether angular averages over $\phi $ make sense for $k$ near $k_{\rm max}$ if only wavevectors $\vk $ within the first BZ are physical
since those averages must also include
wavevectors outside the first BZ.  As discussed above, however, $\vk $ points outside the first BZ can be translated to $\vk $ points within the first BZ using the 
periodic boundary conditions of the space tiled with $O_n'(\vk )$.

Experiments can tune the OAM by changing the splitting of the magnon bands 
with energies $\hbar \omega_{1,2} (\vk )/3JS = 1+\kappa \mp  \eta_{\vk} $ from Eqs.\,(\ref{om1}) and (\ref{om2}) and
\begin{equation}
\eta_{\vk } = \sqrt{\vert \Gamma_{\vk }\vert^2 +d^2\,\Theta_{\vk }^2}.
\end{equation}
Plotted in Fig.\,6, $\eta_{\vk } $ has minima of $0$ at the sides of the BZ for $d=0$ or of $1/3$  
at the midpoints of the sides for $d=0.1$;  
$\eta_{\vk }$ has an absolute maximum of $1$ independent of $d$ at $\vk =0$ and a relative maximum of $3\sqrt{3}\,d\approx 0.51$ for $d=0.1$ at the corners of the BZ,
where $\Gamma_{\vk }=0$.
Therefore, searches for OAM in FM honeycomb materials with significant DM interactions
should concentrate at wavevectors with amplitude $k = 0.385 (2\pi )/a $, where the splitting between magnon bands is approximately $18\sqrt{3}\,dJS=12\sqrt{3}\vert D\vert S$, independent
of the exchange $J$.  Note that the splitting between the upper and lower magnon bands is due to the broken time-reversal symmetry produced by the DM interaction.

\begin{figure}
\begin{center}
\includegraphics[width=8.5cm]{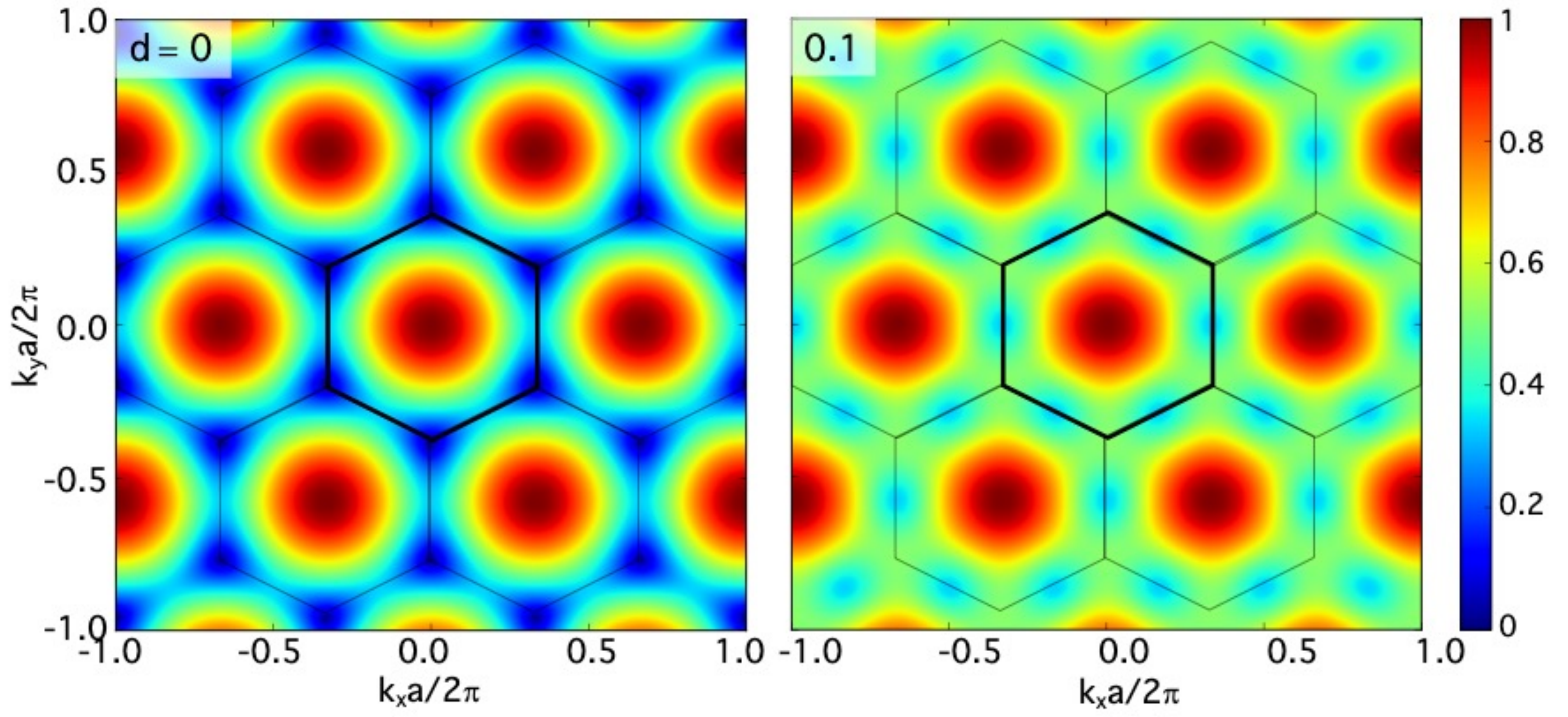}
\end{center}
\caption{The splitting $\eta_{\vk }$ between magnon bands $n=1$ and $n=2$ of the FM open honeycomb lattice with $d=0$ (left) or $0.1$ (right).
The first BZ is shown by the solid hexagon in both figures.}
\label{Fig6}
\end{figure}

If a high-energy electron beam \cite{Mendis22} with transverse momentum 
$ka/2\pi \approx 0.38$ and OAM $L_z=-\hbar $ strikes  a FM honeycomb
sample (like CrI$_3$ \cite{McGuire15, Chen18t} or CrCl$_3$ \cite{Li22}) with $d \approx 0.1$, then Fig.\,4(a) predicts that it is likely to
encounter a magnon with $ka/2\pi \approx 0.38$ and opposite OAM $L_z=\hbar $. 
But the wavevector orientation $\phi $ of that magnon is not determined.  Keep in mind, though, that that the function $F_n(k)$ must be averaged over 
the radial spread $\Delta k$ of the magnon wavepacket \cite{Chang96}.

Several important questions remain unanswered.  Precisely how can $F_n(k)$ be measured?  
In what other systems would $F_n(k)$ be nonzero?  
How do the magnon orbital and spin angular momentum couple to one another?  
We are hopeful that future work will provide answers to these questions as the field of magnonic OAM attracts renewed interest.

Satoshi Okamoto and Giovanni Vignale helped this work with useful conversations.
Research sponsored by the U.S. Department of Energy, 
Office of Science, Basic Energy Sciences, Materials Sciences and Engineering Division.
The data that support the findings of this study are available from the author
upon reasonable request.

\vfill\eject

\end{document}